\documentstyle[preprint,eqsecnum,aps,epsf]{revtex}

\newcommand{\hs}{\hspace*{0.25in}}

\def\decayright#1{\kern#1em\raise1.1ex\hbox{$|$}\kern-.5em\rightarrow}

\begin{document}
\draft
\tighten

\preprint{\parbox{2in}{MADPH-98-1055\\ UCLA-HEP-98-001}}
\title{Tau Neutrino Appearance with a 1000 Megaparsec Baseline}

\author{F. Halzen}
\address{Department of Physics, University of Wisconsin,
Madison, WI 53706}
\author{D. Saltzberg}
\address{Department of Physics and Astronomy,
UCLA,
Los Angeles, CA 90095}

\date{\today}

\maketitle

\begin{abstract}

A high-energy neutrino telescope, such as the operating AMANDA detector, may  
detect neutrinos produced in sources, possibly active galactic nuclei or
gamma-ray bursts, distant by a thousand megaparsecs.  These sources produce
mostly $\nu_e$
or $\nu_\mu$ neutrinos.  Above 1~PeV, $\nu_e$ and $\nu_\mu$ are absorbed
by charged-current interactions in the Earth before reaching the
opposite surface.  However, the Earth never becomes opaque to $\nu_\tau$ since
the $\tau^-$ produced in a charged-current $\nu_\tau$ interaction
decays back into $\nu_\tau$ before losing significant energy.  This
preferential penetration of tau neutrinos through the Earth above $10^{14}$~eV
provides an experimental signature for neutrino oscillations.
The appearance of a $\nu_{\tau}$ component would be evident
as a flat zenith angle dependence of a source intensity at the highest
neutrino energies.  Such an angular dependence 
would indicate $\nu_\tau$ mixing with a sensitivity to
$\Delta m^2$ as low as $10^{-17}$~eV$^2$, for the
farthest sources.  In addition, the presence
of tau neutrino mixing would provide the opportunity for neutrino
astronomy well beyond the PeV cutoff, possibly out to the energies matching
those of  the highest energy protons observed above $10^{20}$~eV.

\end{abstract}

%\begin{center}
%{\it Submitted to Physical Review Letters}
%\end{center}

% 14.60Pq= Neutrino mass and mixing
% 95.85.Ry=Neutrino, muon, pion, and other elementary particles; cosmic rays
% 96.40.Tv=Neutrinos and muons
\newpage
\narrowtext

\section{Introduction}

\hs High-energy neutrino detectors roughly two orders of magnitude larger
in effective telescope area than the Super-Kamiokande experiment are being
constructed to detect astronomical neutrino sources beyond the
Sun~\cite{pr}. 
The most powerful sources may be far beyond the boundaries of our
galaxy, with active galactic nuclei (AGN) and gamma-ray bursts (GRB)
being the leading candidates simply because they are the sources of the
highest energy photons. They may also be the accelerators of the
highest energy cosmic rays.

\hs If AGN and GRBs are the source of the high-energy cosmic-ray spectrum,
which is known to extend beyond $10^{20}$~eV, they will likely produce
neutrinos from the decay of charged pions.  These are the secondary
particles in the interactions of accelerated protons with the very
high density of photons in the source.  Production occurs near the
$\Delta$~resonance in the p-$\gamma$ interaction and the beam is
exclusively composed of $\nu_e$ and $\nu_\mu$.  The flux of neutrinos
is calculable because the properties of the beam and target can be
deduced from the observations of high-energy protons and gamma rays at Earth.  The
prediction is of the order of 50 detected neutrinos per year in a
high energy neutrino telescope with an effective area of
1~km$^2$~\cite{pr,waxman}. Their energies cluster in the vicinity of 100~TeV
for GRBs and 100~PeV for neutrinos originating in AGN jets. For the latter,
even larger fluxes of lower energy energy neutrinos may emanate from
their associated accretion disks~\cite{stecker}.

\hs Whereas AGN exist within 100~Mpc, the most powerful are at
cosmological distances. So are GRBs and therefore, from a particle
physics point of view, we have the extraordinary opportunity to
observe neutrinos which have traveled more than 1000~Mpc.  With this
baseline, the presence of tau neutrinos from such sources would
indicate the presence of neutrino oscillations with $\Delta m^2$ as
low as $10^{-17}$~eV$^2$, where the mass difference squared is
relative to the original $\nu_{e,\mu}$.  For example, such a search
for $\nu_\tau$ appearance
would extend the current searches for $\nu_{\mu}$-$\nu_{\tau}$ oscillations
using atmospheric neutrinos by 14 orders of magnitude.

\hs This particle physics experiment is made possible by the fact that the
$\nu_{\tau}$ can be identified by two of signatures: by
``double-bang'' events from the production and decay of the
$\tau$~lepton, and by the absence of absorption by the Earth.  The
``double-bang'' signature has been described elsewhere~\cite{pakvasa}.
Observation of double-bang events is
difficult in a first generation telescope such as AMANDA.

\begin{figure}[p]
\begin{center}
\leavevmode
\epsfxsize=6.5in
\epsfbox{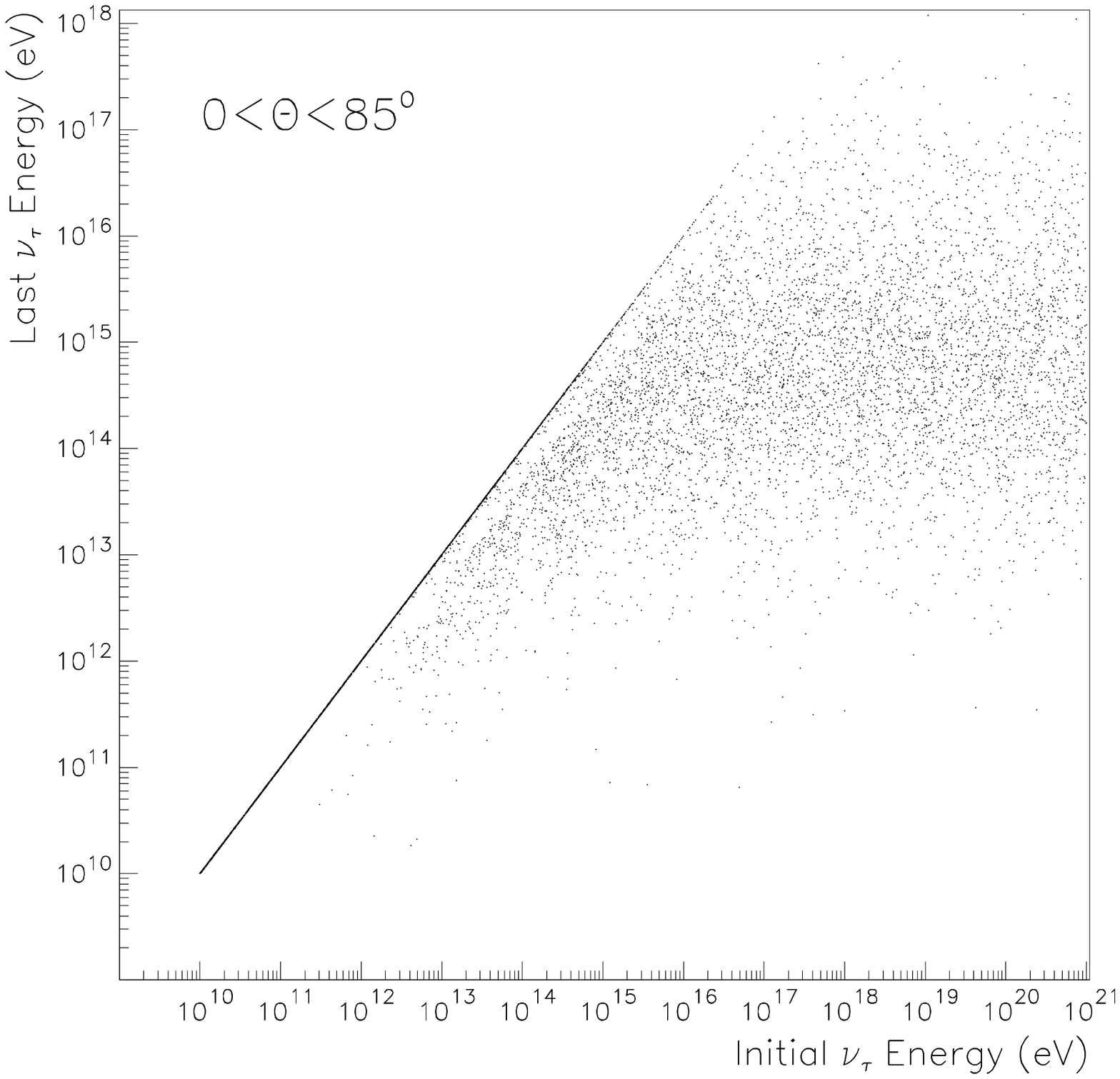}
\caption{Plot of the energy of the final $\nu_\tau$ in the cascade
through the Earth versus the energy of the initial $\nu_\tau$.
The straight line corresponds to neutrinos that do not interact.
Angles larger than $85^{O}$ with respect to
the nadir are excluded because the
$\nu_\tau$ have not yet been moderated to 10$^{15}$~eV.  Note that
an especially hard $\nu_\tau$ input spectrum was chosen for this figure to 
illustrate the effect.}
\label{nuin_nuout}
\end{center}
\end{figure}

\hs We present in this letter a second signature for $\nu_{\tau}$
appearance in a cosmic beam. For energies above 10--100~TeV, $\nu_{e,\mu}$  
neutrinos no longer efficiently penetrate the Earth and
are preferentially observed near the horizon where
they traverse a reduced chord of the Earth.  Our critical observation
is that above these energies the Earth remains effectively transparent to
$\nu_{\tau}$.  A high-energy $\nu_{\tau}$ will interact with the Earth
and produce another $\nu_{\tau}$ of lower energy. In a neutral-current
interaction its energy is, on average, about half. In a charged-current
interaction a $\tau$ is produced which decays in a number of
ways, yet there is always another $\nu_{\tau}$ in the final state.
Its energy is reduced, on average, to about one fifth.
So, high-energy $\nu_{\tau}$'s will initiate a cascade in
the Earth which will contain a $\nu_{\tau}$ of reduced energy in each interaction.  
Once its energy falls below threshold for absorption,
{\it i.e.}, its interaction length becomes comparable to the diameter
of the Earth, the neutrino will propagate to the detector with an
energy in the vicinity of 10--100~TeV.
Figure~\ref{nuin_nuout} shows
the characteristic relationship between the incoming $\nu_\tau$ energy and the
energy of the final $\nu_\tau$ from a simple Monte Carlo.
The simulation includes the $Q^{2}$ dependence of the $W$ propagator and
proton structure~\cite{pr}. The $\nu_\tau$ will be detected
by the appearance of a $\tau^-$ which decays to $\mu^-$ just below the
detector.

\begin{figure}[p]
\begin{center}
\leavevmode
\epsfxsize=6.5in
\epsfbox{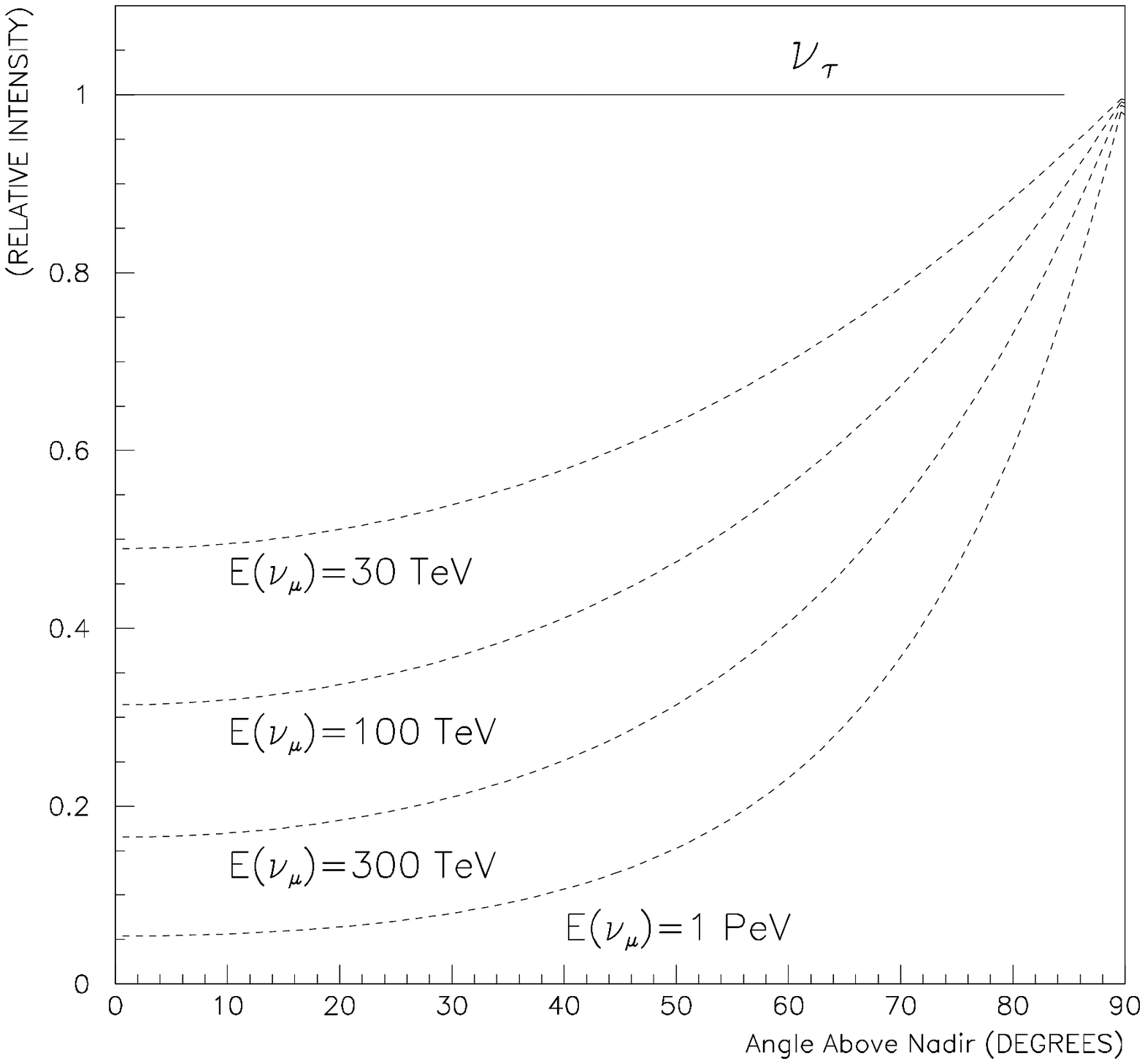}
\caption{Plot of the transmission of $\nu_\mu$ and
$\nu_\tau$ through the Earth's.  The transmission of $\nu_\tau$ is
essentially independent of their energy, as described in the text.
The event rates are normalized to the maximum.}
\label{atten}
\end{center}
\end{figure}

\hs Since all $\nu_\tau$ with energy greater than 100~TeV will
have their energy reduced by this process to about 100~TeV by
the time they reach the detector,
a pile-up of events near 100~TeV would be one clear signature of a
$\nu_{\tau}$ component in the cosmic beam.  However, this would
require the energy spectrum to fall more slowly than $E^{-2}$.  For
diffuse or point sources, the $\nu_{e,\mu}$ will also show a characteristic
absorption as a function of the zenith angle of the
source, but $\nu_{\tau}$ will show none.  It is straightforward to calculate
absorption quantitatively because the neutrino cross sections are 
determined from parton distribution functions that are
constrained by accelerator data.  
Deviation from the
calculated zenith angle distribution towards flatness is a signature
for $\nu_{\tau}$ appearance.  Figure~\ref{atten} compares the zenith
angle distributions of observed source intensities for
$\nu_\mu$ and $\nu_\tau$.

\hs Despite the loss of energy in the cascade, the energies involved are
high enough so that the final $\nu_\tau$ still points back to its
source.  So if there is any large-angle neutrino oscillations involving
$\nu_\tau$ with sufficient $\Delta m^{2}$ (such as that consistent
with the Super-Kamiokande data~\cite{superk})
even a pure $\nu_\mu$ source can be seen
above $10^{15}$~eV, even out to $10^{21}$~eV where the highest energy
cosmic-ray protons have been observed.  These neutrinos will point back
to their sources.

\section{ $\nu_{\tau}$ from gamma-ray bursts: an example}

\hs Recently, GRBs may have become the best motivated source for high
energy neutrinos~\cite{waxman}. Their neutrino flux can be calculated in a
relatively model-independent way.  Although their neutrinos may be
less copious and less energetic than those anticipated from AGN, the
predicted fluxes can be bracketed with more confidence. In GRBs a
fraction of a solar mass of energy ($\sim10^{53}$~ergs) is released
over a time scale of order 1 second as photons with a very hard
spectrum. It has been suggested~\cite{waxmanprime} that, though unknown, the
same cataclysmic events also produce the highest energy cosmic
rays. This association is reinforced by more than the phenomenal
energy and luminosity:

\begin{itemize}

\item
Both GRBs and the highest energy cosmic rays are produced in cosmological
sources, {\it i.e.}, distributed throughout the Universe.

\item
The average rate $\dot E \simeq 4\times10^{44}\rm~Mpc^{-3}~yr^{-1}$
at which energy is injected into the Universe as gamma rays from GRBs
is similar to the rate at which energy must be injected in the
highest energy cosmic rays in order to produce the observed cosmic ray
flux beyond the ``ankle'' in the spectrum at $10^{19}$~eV.
~\\
\end{itemize}

\hs There is increasing observational support for a model where an initial
event involving neutron stars or black holes deposits a solar mass of
energy into a radius of order 100~km~\cite{meegan}. Such a state is opaque
to light. The observed gamma ray display is the result of a
relativistic shock which expands the original fireball by a factor
$10^6$ over 1~second. Gamma rays are produced by synchrotron radiation by
relativistic
electrons accelerated in the shock, possibly followed by
inverse-Compton scattering. The association of cosmic rays with GRBs
obviously requires that kinetic energy in the shock is converted into the
acceleration of protons as well as electrons. It is assumed that the
efficiency with which kinetic energy is converted to accelerated protons is
comparable to that for electrons.  The production of
high-energy neutrinos is a feature of the fireball model because the protons  
will photoproduce pions and, therefore, neutrinos on the gamma rays in the  
burst. We have
a beam dump
configuration where both the beam and target are constrained by
observation: the cosmic ray beam and the observed GRB photon fluxes at
Earth, respectively.

\hs The predicted neutrino flux is~\cite{waxman}
\begin{eqnarray}
	dN/dE &=& A / E^2\qquad            {\rm for}\  E > E_b\\
	      &=& A / (E_b E) \qquad    {\rm for}\   E < E_b
\end{eqnarray}
with $A = 4 \times 10^{-11}$~TeV~(cm$^2$ s sr)$^{-1}$ and $E_b \simeq 100$~TeV.

From an observational point of view, the predicted flux is better summarized  
in terms of the main ingredients of the model:
\begin{equation}
N_\nu = 50 \left[ f_\pi\over 20\% \right] \left[ \dot E\over 4\times10^{44}
{\rm\, Mpc^{-3} \, yr^{-1}} \right] \left[ E_\nu\over 100\rm\ TeV
\right]^{-1} \,,
\end{equation}
{\it i.e.},
we expect 50 events in a km$^2$ detector in one year. Here $f_{\pi}$,
assumed to be 20\%, is the efficiency by which proton energy is
converted into the production of pions and $\dot E$ is the total
injection rate into GRBs averaged over volume and time. The energy of
the neutrinos is fixed by the threshold for photoproduction of pions
by the protons on the GRB photons in the shock. The neutrino rate depends  
weakly on their energy: increased energy per neutrino reduces the flux
as $E^{-1}$. As long as the detected neutrino flux does
not fall off too quickly with energy, the presence of $\nu_\tau$ would
be indicated by a pile-up of events at 100~TeV.\\

\hs Interestingly, this flux may be observable in the currently
operating AMANDA detector.  Its effective area for the detection of
100~TeV neutrinos is 0.24 and 0.06~km$^2$ for $\nu_{\mu}$ and $\nu_e$,
respectively~\cite{hill}. Although these results were obtained by Monte
Carlo simulation, the effective area for a $\nu_e$ can be estimated
from the fact that an electromagnetic shower of 100~TeV energy
produces single photoelectron signals in ice over a radius of 250~m.
The effective area for a $\nu_{\mu}$ is larger because the muon has a
range of 10~km (water-equivalent) and produces single photoelectrons
to 100~m from the track by catastrophic energy losses. Because these
spectacular events arrive with the GRB time-stamp of 1 second
precision and with an unmistakable high-energy signature (they are a
factor $10^{3 \sim 4}$ above AMANDA's nominal threshold), no
background rejection is required.  After correcting for the fact that
BATSE photon detectors only report one quarter of the bursts and that
AMANDA has $4\pi$ acceptance for these events, we predict 30 events
per year.  Such events should be observed in coincidence with a
BATSE burst within a 1 second interval about twice per month.  To
sharpen the evidence one could select high-energy events only, for
instance, events where more than 80 and 150 optical modules report in
the trigger. Such events occur at a rate of 1 and 0.05~Hz, with an
efficiency for 100~TeV neutrinos of 0.4 and 0.1, respectively~\cite{hill}. 
They do
not form an irreducible background because reconstruction can confirm
their GRB origin. About once per year a relatively near burst could
produce multiple events in a single second.

\hs Once a GRB beam has been established, the zenith
angle distribution of the sources can be used to search for  $\nu_{\tau}$
appearance as described above.  The technique may also be applied to
a uniform diffuse source.
%Neutrinos of 100~TeV somewhat low in energy for observing
%absorption in the Earth, see Figure...( this is your initial versus final
%energy figure).

\hs There is also the possibility that high-energy gamma rays and
neutrinos are produced when the shock expands into the
interstellar medium. This mechanism has been invoked as the origin of
the delayed high energy gamma rays~\cite{katz}. The fluxes are expected to be
lower, and are produced over minutes, not seconds, making their
identification more challenging because because matching of the photon and
neutrino directions is now required.

\hs It is important to point out that AGN neutrinos are expected to reach
energies a factor one thousand or more higher than those from
GRBs.~\cite{pr} The predicted rates in the present AMANDA detector
range from copious to non-observable, depending on the theoretical model. If  
one associates
AGN with the source of the highest energy cosmic rays, the predicted
flux is, not surprisingly, similar to the one obtained for GRBs~\cite{halzen}.
Neutrinos are mostly produced near the maximum energy of 100~PeV,
rather than 100~TeV. Such a beam would be ideal for searching for tau
neutrinos.

\section{Conclusion}

\hs We have described a property of tau neutrinos which
allows their efficient detection by neutrino detectors above 1~PeV.
The differential attenuation of $\nu_\mu$ versus
$\nu_\tau$ can be used in existing and future large-area neutrino
detectors  such as AMANDA to search for neutrino oscillations.  A
flat azimuthal dependence of point or diffuse sources
could demonstrate $\nu_\tau$ appearance over a
1000~megaparsec baseline at $10^{15}$~eV, corresponding to a sensitivity
in $\Delta m^{2}$ of $10^{-17}$~eV$^{2}$.
In the presence of neutrino oscillations, these high energy neutrinos
may be useful for finding sources of extremely high-energy
neutrinos such as gamma-ray bursts and  active galactic nuclei.

\begin{thebibliography}{99}
\frenchspacing

\bibitem{pr}
For a review, see T.K. Gaisser, F. Halzen and T. Stanev, {\it Phys.\ Rep.} {\bf
258}(3), 173 (1995); R.~Gandhi, C.~Quigg, M.~H.~Reno and I.~Sarcevic, {\it
Astropart. Phys.}, in press (1995).

\bibitem{waxman}
E.~Waxman and J.~N.~Bahcall, astro-ph/97011231, {\it Phys. Rev. Lett.} {\bf
78}, 2292 (1997).

\bibitem{stecker}
F. W. Stecker and M. Salomon, astro-ph/9501064.

\bibitem{pakvasa}
J. G. Learned and Sandip Pakvasa, {\it Astropart.\ Phys.} {\bf 3}, 267 (1995).

\bibitem{superk}
Y. Fukuda {\it et al.}, hep-ex/9803006.

\bibitem{waxmanprime} E.~Waxman, {\it Phys. Rev. Lett.} {\bf 75}, 386 (1995);  
M.~Vietri, {\it Astrophys. J.} {\bf 453}, 883 (1996).

\bibitem{meegan} C.~A.~Meegan, {\it Nature} {\bf 355}, 143 (1992).

\bibitem{hill}
G. Hill, private communication.

\bibitem{katz} J.I. Katz, Astrophys. Journal {\bf 432}, L27 (1994); F.~Halzen  
and G.~Jaczko, Phys. Rev. D {\bf 54}, 2774 (1996); M. Botcher and C. D.
Dermer, astro-ph/9801027.

\bibitem{halzen}
F. Halzen and E. Zas, Astrophys.J. {\bf 488}, 669 (1997)

\end{thebibliography}
\end{document}